\documentclass[prl,twocolumn,a4paper,superscriptaddress,showpacs]{revtex4}
\usepackage{graphicx,amsmath,bm,natbib, dsfont}

\usepackage[frenchb]{babel}

\def\bra#1{\mathinner{\langle{#1}|}}
\def\ket#1{\mathinner{|{#1}\rangle}}
\def\mn#1{\langle #1 \rangle}
\def\prjct#1{\mathinner{|{#1}\rangle}\!\!\mathinner{\langle{#1}|}}
\newcommand{\coh}[2]{\mathinner{|{#1}\rangle}\!\!\mathinner{\langle{#2}|}}

\def\text#1{\textrm{#1}}
\def\id{\mathds{1}}
\def\E{\mathcal{E}}
\def\eq{\begin{equation}}
\def\eeq{\end{equation}}
\def\tr{\text{tr}}

\def\D{\mathcal{D}}

\def\N{\mathcal{N}}
\def\w{ \omega}

\begin{document}
\title{Macroscopic optomechanics from displaced single-photon entanglement}
\date{\today}
\author{Pavel Sekatski}
\affiliation{Group of Applied Physics, University of Geneva, CH-1211 Geneva 4, Switzerland}
\affiliation{Institut for Theoretische Physik, Universitat of Innsbruck, Technikerstr. 25, A-6020 Innsbruck, Austria}
\author{Markus Aspelmeyer}
\affiliation{Vienna Center for Quantum Science and Technology, Faculty of Physics, University of Vienna, Boltzmanngasse 5, 1-1090 Vienna, Austria}
\author{Nicolas Sangouard}
\affiliation{Group of Applied Physics, University of Geneva, CH-1211 Geneva 4, Switzerland}
\begin{abstract}
Displaced single-photon entanglement is a simple form of optical entanglement, obtained by sending a photon on a beamsplitter and subsequently applying a displacement operation. We show that it can generate, through a momentum transfer in the pulsed regime, an optomechanical entangled state involving macroscopically distinct mechanical components, even if the optomechanical system operates in the single-photon weak coupling regime. We discuss the experimental feasibility of this approach and show that it might open up a way for testing unconventional decoherence models. 
\end{abstract}
\maketitle
%%%%%%%%%%%%%%%%%%

\paragraph{Introduction}
Can a macroscopic massive object be in a superposition of two well distinguishable positions? 
It has been argued that such superpositions undergo intrinsic decoherence, e.g. due to a non-linear stochastic classical field \cite{GRW86, Pearle90, Gisin89} or caused by superposition's perturbation of spacetime \cite{Diosi89, Penrose96}. These decoherence mechanisms are different from conventional decoherence that occurs through entanglement with the environment \cite{Zurek03} and that has been nicely demonstrated in \cite{Brune96, Turchette00, Hackermueller04, Deleglise08, Myatt00}. In contrast, testing for unconventional decoherence models requires a combination of large masses and superpositions of states corresponding to well separated positions. Matter-wave interferometry with large clusters \cite{Nimmrichter11} or with submicron particles \cite{Romero-Isart11} is one possible route. Another approach is to manipulate states of motion of massive mechanical resonators, a fast moving field of research that has now succeeded in entering the quantum regime \cite{Oconnell10, Teufel11, Chan11}. In the framework of optically controlled mechanical devices \cite{Aspelmeyer13}, the proposals \cite{Bose97, Bose99} have the potential to create a superposition of mechanical states with a distance of the order of the mechanical zero-point fluctuation where the effects of unconventional decoherences might be observable \cite{Marshall03, Kleckner08}. However, this requires (i) to work in the single-photon strong coupling regime, (ii) a coupling rate at least of the order of the mechanical frequency so that the displacement induced by a single photon is larger than the mechanical zero point spread, (iii) to work in the resolved sideband regime where the mechanical frequency is larger than the cavity decay rate to allow ground state cooling. While (i) and (ii) can be relaxed, e.g. using nested interferometry \cite{Pepper12prl} and (iii) can be circumvented by cooling e.g. via pulsed optomechanical interactions \cite{Vanner11}, the distance between the superposed states remains small, of the order of the mechanical ground state extension.

Here, we show how to create macroscopic optomechanical entanglement with relatively simple ingredients. Our proposal starts with an optical entangled state of the type $\ket{\bar{+}}_A \ket{-}_B - \ket{\bar{-}}_A \ket{+}_B,$ involving two spatial modes $A$ and $B.$  Concretely, this state is obtained by sending a single photon into a beamsplitter (with output modes $A$ and $B$) and by subsequently applying a phase-space displacement on $A$. The displaced photons in $A$ then interact with a mechanical system $M$ through radiation pressure. If the interaction between $A$ and $M$ falls within the pulsed regime \cite{Cerrillo11, Vanner11, Wang11} where the pulse duration is much smaller than the mechanical period, the optical and mechanical modes entangle, $|\bar{+}\rangle_{AM} |-\rangle_B - |\bar{-}\rangle_{AM} |+\rangle_B.$ Because $\ket{\bar{+}}_A$  and $\ket{\bar{-}}_A$ are well distinguishable in photon number, the mechanical components $\rho_M^{(\pm)}=$tr$_A \{|\bar{\pm}\rangle\langle\bar{\pm}|_{AM}\}$ are well distinct in the phase space even in the weak coupling regime and if the coupling rate is smaller than the mechanical frequency. This relaxes the constraints on the initial cooling of the mechanical oscillator and makes our proposal well suited to test unconventional decoherence processes, as we show below.\\

\paragraph{Optomechanical entanglement}
\begin{figure}
\includegraphics[width=0.6\linewidth]{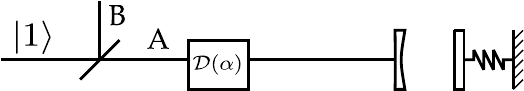} 
\caption{A single photon is sent through a beamsplitter and creates an entangled state between the two output modes $A$ and $B.$ $A$ then undergoes a displacement and couples to a mechanical system by momentum transfer in the pulsed regime.
}
\label{Fig1}
\end{figure}
Consider an optomechanical cavity described by 
$
H=\hbar \omega_m m^\dag m-\hbar g_0 a^\dag a (m^\dag + m)
$
where $\omega_m$ is the angular frequency of the center of mass motion of the mechanical system, $m, \, m^{\dag}$ $(a, \, a^{\dag})$ are the bosonic operators for the phononic (photonic) modes and $g_0=\frac{\omega_c}{L}\sqrt{\frac{\hbar}{2 \text{M} \omega_m}}$ for a Fabry-Perot cavity with a mechanically moving end mirror ($\omega_c$ is the optical angular frequency, $L$ is the cavity length and M is the effective mass of the mechanical mode). The form of the optomechanical interaction, proportional to $a^\dag a \bar x_m$ (where $\bar x_m=x_0 (m+m^{\dag})$ is the position operator, $x_0= \sqrt{\frac{\hbar}{2M \omega_m}}$ being the mechanical zero-point fluctuation amplitude) tells us that starting with a superposition of photonic components that are well distinguishable in photon number space, we can create a superposition of mechanical states corresponding to well distinct momenta. Displaced single-photon entanglement exhibits such a property \cite{Sekatski12, Sekatski13} and has the advantage of being easily prepared, see Fig. \ref{Fig1}. It can be written as
\begin{equation}
\label{photon_state}
\frac{1}{\sqrt{2}}\left(\mathcal{D}(\beta)|+\rangle_A|-\rangle_B - \mathcal{D}(\beta)|-\rangle_A |+\rangle_B\right)
\end{equation}
where $|\pm\rangle=2^{-\frac{1}{2}}\left(|0\rangle \pm |1\rangle\right),$ $|0\rangle$ being the vacuum, $|1\rangle$ the single photon Fock state \cite{foonotesinglet}. $\mathcal{D}(\beta)=e^{\beta a^\dag-\beta^\star a}$ stands for the displacement operator and can be implemented using an unbalanced beamsplitter and a coherent state \cite{Paris96}. Although the photon number distributions for $\mathcal{D}(\beta)|+\rangle_A$ and $\mathcal{D}(\beta)|-\rangle_A$ partially overlap (their variance is given by $\beta^2+\frac{1}{4}$), their mean photon numbers $\beta^2 �\pm \beta+\frac{1}{2}$ are separated by 2$\beta$ \cite{Sekatski12}. (Here $\beta$ is considered real, as all along the paper). In other words, their distance in the photon number space is of the order of the square root of their size. This makes the state (\ref{photon_state}) macroscopic in the sense that its components can be distinguished without a microscopic resolution \cite{Sekatski13}. \\

Consider first the case where $|n\rangle_A$ photons interact with the mechanical mode initially prepared in its motional ground state $|0\rangle_M.$ According to \cite{Bose97}, they induce a coherent displacement of the mechanical state whose amplitude varies periodically in time $e^{i \frac{g_0^2 n^2}{\omega_m^2} (\omega_m t - \sin(\omega_m t))} |\frac{g_0 n}{\omega_m}(1-e^{-i \omega_m t})\rangle_M$ $|n\rangle_A.$ The first exponential term corresponds to the variation of the cavity length and is quadratic in the photon number because the mean position of the mechanical oscillator depends on the photon number. To avoid this non-linear behavior, we consider the pulsed regime where the interaction time $\tau$ is much smaller than the mechanical period ($\sin(\omega_m \tau) \sim \omega_m \tau$, c.f. below for the detailed conditions). Right after this interaction, the propagator has the simple form 
$
e^{ig_0 \tau a^\dag a (m+m^\dag)}
$
and after a free evolution of duration $t,$ the overall propagator can be written as
$
U(t)=e^{ig_0 \tau a^\dag a (e^{i\omega_m t}m+e^{-i\omega_m t}m^\dag)}e^{-i\omega_m t m^\dag m}.
$
An initial state $|0\rangle_M|n\rangle_A$ now evolves towards 
$
|n \alpha(t) \rangle_M |n\rangle_A
$
where  $|n\alpha(t)\rangle_M$ is a coherent state with a fixed amplitude and a periodic phase $n\alpha(t)= - i g_0 n \tau e^{-i\omega_m t }.$ In other words, the $n$ photons kick the mechanical mode that gets an additional momentum $2 g_0 n \tau p_0$ at time $t=0$ ($p_0=\sqrt{\frac{\hbar M \omega_m}{2}}$ is the initial mechanical momentum spread). The mechanical state then starts to rotate in phase space. It reaches a minimal position $-2 g_0 n \tau x_0$ after $\frac{\pi}{2\omega_m},$ then gets a momentum $-2 g_0 n \tau p_0$ after $\frac{\pi}{\omega_m}$ and so on. \\
\begin{figure}
\includegraphics[width=0.7\linewidth]{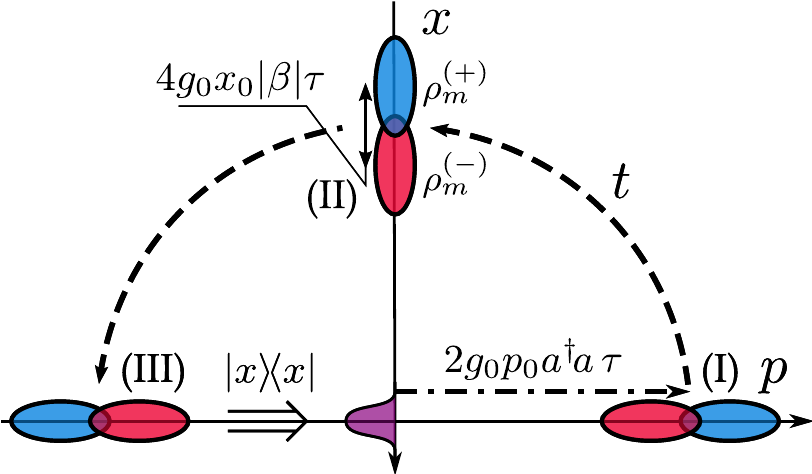}
\caption{Trajectory of the mechanical state in the phase space. (I) The mirror first gets a momentum proportional to the mean photon number. The superposition of two mechanical states (red and blue ovals) result from the interaction with a superposition of $\D(\beta)\ket{-}_A$ and $\D (\beta) \ket{+}_A$. (II) After a quarter of a period, the positions of the two superposed states are maximally distinct and are correlated with the $\bar X$ quadrature of the mode B. (III) By measuring the position after a multiple of a half period, the information about the number of photons in $A$ (contained in the mirror) is erased, which enable to observe the entanglement between $A$ and $B$.
}
\label{Fig2}
\end{figure}
Let us now come back to the initial state (\ref{photon_state}). The pulse in $A$ enters the optomechnical cavity, the mechanical mode being in $\ket{0}_M$, as before. A time $t$ after the interaction, the state of the system is 
\begin{eqnarray}
\nonumber
\frac{1}{\sqrt{2}}&&
\Big( \sum_{k} {a_\beta^{(+)}(k)} |k\rangle_A |k\alpha(t)\rangle_M |-\rangle_B -\\
 &&\sum_k {a_\beta^{(-)}(k)} |k\rangle_A |k\alpha(t)\rangle_M |+\rangle_B  \Big)
\label{coupled_state_pm}
\end{eqnarray}
where $a_\beta^{(\pm)}(k)=\frac{1}{\sqrt{2}}e^{-\frac{\beta^2}{2}}\frac{\beta^{k}}{\sqrt{k!}} \left(1 \pm \left(\frac{k}{\beta}-\beta\right)\right)$ are the probability amplitudes for having $k$ photons in $\mathcal{D}(\beta)|\pm\rangle_A.$ 
Since $\sum_k \big(a_\beta^{(+)}(k)\big)^\star a_\beta^{(-)}(k)=0,$ the mechanical mode entangles with the optical modes. Specifically, after $\frac{\pi}{2\omega_m},$ the state (\ref{coupled_state_pm}) involves two mechanical states $\rho_M^{(\pm)}=\sum_{k} |a_\beta^{(\pm)}(k)|^2  |\text{-}g_0\tau k \rangle_M\langle \text{-}g_0\tau k |,$ each having a variance $\left(1+g_0^2 \tau^2 \left(1+4 \beta^2\right)\right) x_0^2$ in space and for which the mean position is separated by $4 g_0 \tau \beta x_0$ (see fig. \ref{Fig2}). These two mechanical states can thus be distinguished with a detector having a resolution $\delta x \sim 2 g_0 \tau \beta x_0,$ see below. For $g_0 \tau \beta \geq 1,$ such a detector cannot resolve two phononic Fock states with $n$ and $n+1$ excitations  (no microscopic resolution) and the entangled state (\ref{coupled_state_pm}) can fairly be defined as being macroscopic.\\

\paragraph{Macroscopic correlations}
We now show how to demonstrate that the mechanical mode involves macroscopically distinct states $\rho_M^{(\pm)}$. 
\begin{figure}
\includegraphics[width=0.6\linewidth]{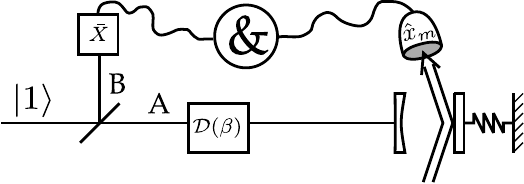}
\caption{Setup for checking that the result of the homodyne measurement on $B$ is correlated with the position of $M$ even if the position measurement does not have a microscopic resolution.}
\label{Fig3}
\end{figure}
More precisely, we show that $B$ and $M$ are correlated, i.e. when the state of B is projected into $|-\rangle$ ($|+\rangle$), the mechanical mode is found in $\rho_M^{(+)}$ ($\rho_M^{(-)}$) a quarter of a mechanical period after the interaction, (c.f. Fig. \ref{Fig3}) and that these correlations can be revealed without the need for a microscopic resolution. This is done by tracing out $A,$ and by measuring the $\bar X=2^{-1/2}(b^{\dag}+b)$ quadrature of $B$ and the mirror position. The latter can be realized following \cite{Vanner11}, by observing through a quadrature measurement the phase acquired by a strong, short light pulse reflected by the mechanical oscillator. We attribute the value $+1$ ($-1$) to a positive (negative) result of the quadrature measurement on $B$ and $+1$ $(-1)$ if the mirror is found to be shifted more to the left (right) with respect to its mean position $-g_0 \tau x_0(1+2 \beta^2)$. For an uncertainty $\delta x$ on the measurement of the mirror position, the probability $P_{\pm E_\pm}$ for having the same results $\{\pm1,\pm1\}$ is given by $\frac{1}{4}+\frac{g_0\tau \beta}{2\pi\sqrt{1+g_0^2\tau^2 \beta^2+\delta x^2/(4x_0^2)}}$ (for $\beta \gg 1$) while the probability for having different results $P_{\pm E_\mp}=\frac{1}{2}-P_{\pm E_\pm}.$ Therefore, the correlations between the outcomes (the probability for having correlated results minus the probability for having anti-correlated results) are given by 
$
\frac{2}{\pi} \frac{g_0 \tau \beta }{\sqrt{1+g_0^2 \tau^2 \beta^2+\delta x^2/(4x_0^2)}}.
$
In the regime of interest $g_0 \tau \beta \geq 1,$ even a coarse grained measurement with the resolution $\delta x = 2g_0 \tau  \beta x_0$ leads to substantial correlations $\sim 0.45.$ This is a consequence of the macroscopic characteristic of the optomechanical state (\ref{coupled_state_pm}).\\

\begin{figure}
\includegraphics[width=0.6\linewidth]{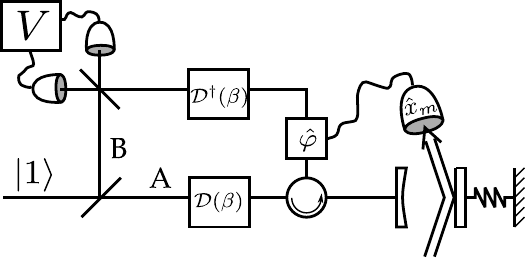}
\caption{Setup for probing the effect of decoherence on the interference between $A$ and $B$ obtained after erasing the "which path" information gained by the mirror. A feedback loop is needed to control the path length of $A$ (relative to $B$) depending on the result of the measurement of the mirror position.
}
\label{Fig4}
\end{figure}

\paragraph{Testing unconventional decoherence models}
Fig. \ref{Fig4} shows how to probe the effect of mirror decoherence. First, the mechanical position is measured at any time that is a multiple of half a mechanical period where no information is obtained about the state of $A.$ Finding the mirror at the position $y$ projects the overall state into
\begin{eqnarray}
\nonumber
\frac{1}{\sqrt{2}}&&
\Big( \sum_{k} a_\beta^{(+)}(k) e^{i\sqrt{2} g_0 \tau k y} |k \rangle_A |-\rangle_B -\\
 &&\sum_k a_\beta^{(-)}(k) e^{i\sqrt{2} g_0 \tau k y} |k\rangle_A |+\rangle_B \Big) |y\rangle_M.
\label{coupled_state}
\end{eqnarray}
Actively controlling the relative length of paths $A$ and $B$ to get rid of the undesired phase term 
$e^{i\sqrt{2} g_0 \tau k y}$ and subsequently applying $\mathcal{D}(-\beta)$ leaves 
the optomechanical state in 
$
\frac{1}{\sqrt{2}}\left(|1\rangle_A|0\rangle_B - |0\rangle_A |1\rangle_B\right)|y\rangle_M.
$
The modes $A$ and $B$ can then be combined on a beamsplitter and varying their relative phase leads to interference fringes, ideally with a unit visibility $(V).$ (Note here that from the values of the probabilities $p_{mn}$ of detecting $m\in\{0,1\}$ photons in A and $n\in\{0,1\}$ in B, a lower bound on the negativity between $A$ and $B$ can be obtained $\N_{AB} \geq \frac{1}{2}\big(\sqrt{(p_{00}-p_{11})^2 + ( V (p_{01}+p_{10}))^2} - (p_{00}+p_{11})\big)$ through the approach presented in \cite{Chou05}.) Decoherence of the mirror operates as a weak measurement of the photon number on $A$ (see SM). Therefore, if the measurement of the mechanical position is delayed, more and more "which path" information is revealed which decreases the visibility as the delay time increases. Importantly, different types of decoherence give rise to different behaviors of the visibility as the delay time increases. As particular examples, we compare conventional (environmental induced) decoherence with unconventional decoherence proposed by gravitationally induced collapse \cite{Diosi89, Penrose96} and by quantum gravity \cite{Ellis84}  (see SM). For sufficiently large $\beta,$ i.e. macroscopic entanglement, and small thermal dissipation we find an experimentally feasible parameter regime, in which the unconventional decoherence rates surpass the conventional ones, hence opening up the possibility for experimental tests (see below). Finally, note that the observed visibility is degraded if the mirror position is not accurately measured. A small imprecision $\delta x$ would indeed introduce an additional phase on $A$ that prevents its re-displacement to the single photon level and degrades the quality of the interference between $A$ and $B$ \cite{Sekatski12, footnote_phase_dis}. Quantitatively,  
\begin{equation}
\label{visibility}
V \approx 1- \frac{3}{2}\delta \phi^4 \beta^4 + o(\delta \phi^4 \beta^4)
\end{equation}
where $\delta \phi = \frac{\delta x}{x_0}  \big(2 \big(\frac{\delta x}{x_0}\big)^2 +1\big)^{-\frac{1}{2}} \sqrt{2} g_0 \tau.$
A high accuracy $\delta x \lesssim \frac{x_0}{(g_0 \tau \beta)^2}$ is thus required to observe high visibility and to see the effect of mirror decoherence. \\

\paragraph{Witnessing optomechanical entanglement} We can prove that the mirror is entangled with the optical modes from an entanglement witness which uses the values of $\{ P_{\pm E_\pm}, P_{\pm E_\mp} \}$ and $\N_{AB}$ only (see SM). The witness is based on the following intuitive argument: since $B$ is a qubit, the only way for $M$ to be correlated to $B$ and for $B$ to be entangled with the joint system $AM$ is that $M$ is entangled with $AB.$ Concretely, we can conclude about optomechanical entanglement if $\N_{AB} > \sqrt{P_{+ E_+} P_{- E_+}}+\sqrt{P_{+ E_-}P_{- E_-}}.$ We emphasize that in contrast to the correlation measurement, the detection of entanglement $\N_{AB}$ requires a measurement of the mirror position with a very high accuracy (through $V$). We are retrieving what seems to be the essence of macro entangled states: although they involve components that can easily be distinguished without microscopic resolution, one needs detectors with a very high precision to reveal their quantum nature \cite{Sekatski13}.\\

\paragraph{Experimental feasibility}
We now address the question of the experimental feasibility in detail. First, we require 
$
4 g_0 \tau \beta \gtrsim 1
$
which allows one to observe significant correlations between $M$ and $B.$ 

To further guarantee a high visibility of the interference between $A$ and $B,$ the system needs to operate in the linear regime. For a pulsed optomechanical interaction $(\tau \ll \omega_m^{-1}),$ the non-linear response of the optomechanical system degrades the visibility of the interference pattern according to $V \, \rightarrow \, V(1-\epsilon)$ where $\epsilon=(g_0 \tau \beta)^6 \frac{\omega_m^2}{g_0^2 \beta^2}$ \cite{footnote_phase}. This undesired effect is thus negligible if $g_0 \beta \gg \omega_m.$ The requirement of observing a high interference visibility also imposes the mirror position to be accurately estimated, c.f. eq. (\ref{visibility}). It has been established in \cite{Vanner11} that the maximum accuracy is obtained by choosing an input drive with a duration $\sim \frac{\ln 2}{\kappa}.$ The achievable precision then depends on its number of photons $N_p$ via $\frac{\delta x}{x_0} = \frac{\kappa}{\sqrt{5}g_0 \sqrt{N_p}}$ and is thus high if $g_0 \sqrt{N_p} > \kappa.$ The primary limitation for $N_p$ is the power that can be homodyned before photodetection begins to saturate. Assuming a saturation power of 10 mW results in $N_p \sim 5 \times 10^{16}/\kappa.$ To build up a proposal as simple as possible, we consider the case where a single local oscillator with a controllable amplitude is used both for implementing the displacement and for measuring the mirror position ($\tau\sim \frac{\ln 2}{\kappa}$). Using eq.  (\ref{visibility}), this results in the reduced visibility $V\rightarrow \, V(1-\bar \epsilon)$ where $\bar\epsilon \sim 2 \times 10^{-35} \kappa^2 \beta^4.$

The mechanical device also needs to be prepared in its ground state. More precisely, if the mechanical oscillator is initially in a thermal state with a mean occupation $n_{\text{th}},$ the interference visibility is unchanged but the observed correlations decrease according to 
$
\frac{2}{\pi} \frac{g_0 \tau \beta }{\sqrt{n_{\text{th}}+1+g_0^2 \tau^2 \beta^2+\delta x^2/(4x_0^2)}}.
$
High correlations can thus be observed if $\sqrt{n_{\text{th}}} \ll g_0 \tau \beta$ i.e. the constraint on the initial cooling is relaxed for macroscopically distinct mechanical states. Cooling in the pulsed regime can be obtained through various schemes \cite{Vanner11, Cerrillo11}. For example, Refs. \cite{Vanner11, Vanner13} show that two subsequent pulses (identical to the pulses used for the measurement of the mechanical position) that are separated by $\frac{\pi}{2\omega_m}$ allow to cool the mechanical mode to an effective thermal occupation of $n_{\text{eff}} = \frac{1}{2} (\sqrt{1+\frac{\kappa^4}{g_0^4 N_p^2}}-1).$ For $g_0 \sqrt{N_p} > \kappa,$ this results in $n_{\text{eff}} \ll 1,$ i.e. ground state cooling. 

For concreteness, we consider a mechanical mirror with resonance frequency $\omega_m=2\pi�\times 20 \times 10^3$ s$^{-1}$ ($\frac{2\pi}{\omega_m}=50$ $\mu$s) and an effective mass M$=60$ ng in a $0.5$ cm long cavity ($g_0/\omega_m=5\times 10^{-3}$). We require correlations larger than 0.5 ($4 g_0 \tau \beta=6$) and an error on the overall visibility of $\sim 1\%.$ This imposes a cavity finesse of $\sim 8000$ ($\beta \sim 40000,$ $N_p \sim 4 \times 10^9,$ $\kappa \sim 2\pi \times 2 \times 10^6$ s$^{-1},$ $\tau \sim 60$ ns). For comparison, the highest reported finesse in an optical Perot-Fabry cavity with micromirrors is $1.5 \times 10^5$ \cite{Muller10}. 

The photons in $A$ need to be stored on the timescale of the decoherence being probed. A simple fiber loop allows one to reach delay times up to 100 $\mu$s without significant loss at telecom wavelength. Much longer delays can be obtained with such a technique if one is willing to use postselections \cite{footnotepostselection}.

The surrounding temperature $T$ must also be low enough so that the effect of conventional (environmentally induced) decoherence \cite{Zurek03} is negligible on the timescale of the decoherence being probed. This requires $T \ll \frac{\hbar  \omega_m Q_m}{k_B}\frac{1}{2(g_0 \tau)^2 \beta^2}{\frac{1}{2\pi n}}$ for $n$ mechanical periods.  In other words, for a base temperature of $T=800$ mK and a mechanical quality factor of $Q_m=10^6,$ conventional  decoherence operates on a timescale of 1 $\mu$s, which is long enough to observe optomechanical entanglement. Lower temperatures and/or higher $Q_m$ are required for testing unconventional decoherence models.  For example, for quantum gravity induced collapse \cite{Ellis84}, we find a timescale $\sim 415$ $\mu$s following \cite{Pepper12},  which would be testable with the proposed device with $Q_m \sim 1.5 \times 10^7$ and $T \sim 20$ mK where conventional decoherence operates on $\sim 630$ $\mu$s. % Let us mention that commercially available silicon nitride membranes have been demonstrated with $Q_m$ up to $10^7$ in cavities with finesse of over $10^4$ \cite{Zwickl08}. 
Gravitationally induced decoherence \cite{Diosi89, Penrose96} provides another example, despite the known ambiguity with respect to the mass distributions. Under the assumption where the mass is distributed over spheres corresponding to the size of atomic nuclei, we find a timescale of $\sim 10$ $\mu$s following \cite{Kleckner08}. This is testable with the proposed device for $T \sim 300$ mK and $Q_m \sim 10^7$ where conventional decoherence operates on $\sim 30 \, \mu$s. Note that in addition to absolute decoherence rates, the scaling behavior with respect mechanical parameters e.g. the mass, provides an independent assessment of the nature of the observed decoherence (see SM).\\

\paragraph{Conclusion}%__________________________________________________________________________________
 We have proposed a way for creating and detecting macroscopic optomechanical entanglement which combines displaced single-photon entanglement and pulsed optomechanical interaction. Our proposal can be implemented in a wide variety of systems. The optomechanical photonic crystal cavity device introduces in Ref. \cite{Safavi13} could exhibit correlations of 0.6 and an interference visibility of 0.95 at a temperature of a few kelvins, while more massive systems, like the one proposed before, open up a way to measure unconventional decoherence models.\\

\paragraph{Note added} During the completion of this work, we became aware of a related work by Ghobadi and co-workers. Our results have been jointly submitted to Physical Review Letters in August, the 30th 2013.\\

\paragraph{Acknowledgements} We thank N. Gisin, K. Hammerer, S. Hofer, N. Timoney, and P. Treutlein for discussions. This work was supported by the Swiss NCCR QSIT, by the Austrian Science Fund FWF (SFB FoQuS), the Vienna Science and Thechnology Fund WWTF, the European Research Council ERC (StG QOM) and the European Commission (IP SIQS, ITN cQOM).\\

%%%%%%%%%%%%%%%%%%
\textbf{Supplemental Materials}\\
\\
\\
\centerline{\textit{Witnessing optomechnical entanglement}}\\

We here show rigorously that the measurement of correlations $\{ P_{\pm E_{\pm}}, P_{\mp E_{\pm}} \}$ between B and M and the lower bound on the negativity $\N_{AB}$ between A and B suffice to witness optomecanical entanglement, i.e. to prove that the state of the global system is not separable with respect to the partition AB$|$M. \\

Let us start by assuming the opposite, i.e. $\rho_{AB|M}=\sum_i\, p_i  \rho_{AB}^i \otimes \rho_M^i.$ The correlations between AB and M being classical, they can be unlimitedly broadcasted. In particular, one can write 
\eq
\rho_{AB|M}=\tr_E \underbrace{\sum_i p_i\,  \rho_{AB}^i \otimes \rho_M^i\otimes \rho_E^i}_{\rho_{ABME}}
\eeq
with $\rho_M^i=\rho_E^i$ and thus $\rho_{AB|M}=\rho_{AB|E}$ (with $\rho_{AB|E}=\tr_M \sum_i p_i\,  \rho_{AB}^i \otimes \rho_M^i\otimes \rho_E^i)$. In other words, the environment $E$ has the same information about the state of the optical system AB than the mirror M. \\

Now, let us focus on the entanglement remaining between the qubit B and AM once the environment is traced out $\E_{B|AM}(\tr_E \rho_{ABME}).$ Here $\E$ is an arbitrary convex entanglement metric. For any POVM $\{E_j\}$ operating on the environment $\sum_j E_j =\id$
\eq
\E_{B|AM}(\tr_E \rho_{ABME}) \leq \sum_j \E_{B|AM}( \underbrace{\tr_E E_j \rho_{ABME}}_{c_j \rho_{ABM}^j}).
\eeq 
The entanglement (between B and AM) in every state $\rho_{ABM}^j$ is bounded by the fact that the measurement outcome $E_j$ provides some knowledge on the partial state of the qubit B. In particular, if the states of E and B are perfectly correlated, the states $\rho_{ABM}^j$ are separable with respect to the partition B$|$AM. %(This holds because B is a qubit: a qubit can only be maximally entanglement with another system iff its partial state is maximally mixed). 
 More concretely, let us consider the decomposition of each state $\rho_{ABM}^j$ as a statistical mixture of pure state $\ket{\psi_k^j}$ i.e.  $\rho_{AMB}^j=\sum_k \lambda_k \prjct{\psi_k^j}$ with $\sum_k \lambda_k=1.$ We have 
$$
\E_{B|AM}(\rho_{AMB}^j)\leq \sum_k \lambda_k \E_{B|AM}(\prjct{\psi_k^j}).
$$ 
Furthermore, any pure state $\ket{\psi_k^j}$ can be written as $\sqrt{p_+^{j,k}}\ket{+}_B\ket{\xi_+^{j,k}}_{AM}+\sqrt{p_-^{j,k}}\ket{-}_B\ket{\xi_-^{j,k}}_{AM}$ (because B is a qubit). Further consider a particular convex measure, the negativity $\N.$ We have $\N_{B|AM}(\psi_k^j)\leq \sqrt{p_+^{j,k} p_-^{j,k}} $ and from the Cauchy-Schwarz inequality
\begin{eqnarray}
\nonumber
&\N_{B|AM}(\rho_{ABM}^j)& \leq \sum_k \lambda_k  \sqrt{p_+^{j,k} p_-^{jk}}  \\
\nonumber
&&\leq  \big( (\sum_k \lambda_k   p_+^{j,k}) (\sum_ k \lambda_k p_-^{j,k})\big)^{1/2} \\
\nonumber
&&
=  \sqrt{P_{+|j} P_{-|j} }
\end{eqnarray}
%for which the concurrence (we explicit the entanglement measure now) $\N_{B|AM}(\psi_k^j)\leq \sqrt{p_+^{j,k} p_-^{j,k}} $. By Cauchy-Schwarz inequality $\sum_k \lambda_k  \sqrt{p_+^{j,k} p_-^{jk}}  \leq  \big( (\sum_k \lambda_k   p_+^{j,k}) (\sum_ k \lambda_k p_-^{j,k})\big)^{1/2} =  \sqrt{P_{+|j} P_{-|j} }$, 
where $P_{+|j} = \tr \prjct{+}_B \rho_{ABM}^j$ and $P_{-|j} = \tr \prjct{-}_B \rho_{ABM}^j$ are the marginal probabilities for B. Therefore the negativity of the state $\rho_{AB|M}$ is upper bounded by 
\eq
\N_{B|AM}(\rho_{AB|M})\leq \sum_j  c_j \sqrt{P_{+|j} P_{-|j}}=  \sum_j   \sqrt{ P_{+j} P_{-j} }
\eeq
where 
\begin{eqnarray}
\nonumber
&P_{\pm j}&=c_j P_{\pm |j}\\
\nonumber
&&= \tr\, E_j\!\otimes\!\prjct{\pm}_B \rho_{ABME} \\
\nonumber
&&= \tr\, E_j\!\otimes\!\prjct{\pm}_B \rho_{AB|E} \\
\nonumber
&&=\tr\, E_j\!\otimes\!\prjct{\pm}_B \rho_{AB|M}
\end{eqnarray}
is the probability to simultaneously find the qubit $B$ in the state $\ket{\pm}_B$ and to get the outcome $j$ when applying the POVM $\{E_j\}$ on the mirror (notice that  the POVM now acts on the mirror). \\

Consider the POVM composed with two elements $E_+=\int_{\mn{x}}^\infty dx \prjct{x}$ et $E_- =\int_{-\infty}^{\mn{x}} dx \prjct{x} $ where $\mn{x}= - g_0 \tau x_0 (1 + 2 \beta^2)$ is the mean mirror position. In this case, the probabilities $P_{+j}$ correspond exactly to the probabilities $P_{+E_i}, \-\ i \in \{+,-\}$ introduced in the main text that have been used to show that B and M are correlated even if the measurement of the mirror position does not have a microsocpic resolution. In summarize, if the optomechanical state is separable with respect to the partition AB$|$M, the following inequality is fulfilled 
\eq\label{ent bound}
\N_{B|AM}(\rho_{AB|M})\leq \sqrt{ P_{+E_+} P_{-E_+} } + \sqrt{ P_{+E_-} P_{-E_-}}.
\eeq 
\\

%The last expression characterises the correlations between a pojective measurement on a qubit $B$ and the POVM $\{ E_j \}$ on the mirror. For the threshold position measurement $E_+=\int_{\mn{x}}^\infty dx \prjct{x}$ et $E_- =\int_{-\infty}^{\mn{x}} dx \prjct{x} $ with the mean mirror position $\mn{x}= - g_0 \tau x_0 (1 + 2 \beta^2)$, the negativity is upper bounded by $ 2\sqrt{ P_{+E_+} P_{-E_+} } + 2\sqrt{ P_{+E_-} P_{-E_-} } = o_{BM}$. Specifically for the proposed state $o_{BM}\approx \sqrt{1-\frac{2}{\pi}\big(1- \frac{1}{1+ g_0^2 \tau^2\beta^2}\big)}$, valid  for large $\beta$. 
%Accordingly, the negativity between B and AM is lower bounded by the correlation between $B$ and $E$ (or equivalently between $B$ and $M$).  In summarize
%\eq\label{ent bound}
%\rho_{AB|M}\,\text{separable}  \implies \N_{A|BM}(\rho_{AB|M})\leq o_{BM}.
%\eeq 

We have shown in the main text how to access the negativity between A and B once the mirror position is measured. The idea consists in (i) measuring the mirror position (ii) using a feedback loop to compensate the relative phase between A and B depending on the result of the mirror position and (iii) undoing the displacement of the mode $A$. This results in an entangled state between the optical modes only, each being filled with a small number of photons. We can then measure the quantity of entanglement between A and B using the tomographic approach based on single photon detections presented in \cite{Chou05} and successfully applied in \cite{Laurat07, Choi08, Usmani12, Bruno13}. This approach reveals the entanglement in the subspace where there is at most one photon in each mode. More precisely, the modes A and B are combined on a balanced beam-splitter and varying their relative phase leads to interference fringes with the visibility V. From the values of probabilities $p_{mn}$ of detecting $m\in\{0,1\}$ photons in A and $n\in\{0,1\}$ in B, a lower bound on the negativity between A and B is obtained through 
$$
\N_{AB} \geq \frac{1}{2}\big(\sqrt{(p_{00}-p_{11})^2 + ( V (p_{01}+p_{10}))^2} - (p_{00}+p_{11})\big)
$$ 
\cite{Chou05,Laurat07, Choi08, Usmani12, Bruno13}. Since the steps (i)-(iii) operate locally on AM, the measured negativity $\N_{AB}$ provides a lover bound on the negativity between  $B$ and $AM,$ i.e.
$$
\N_{AB} \leq \N_{B|AM}
$$
Therefore, if the results of measurements are such that
\eq
\label{witn_ent}
\N_{AB} >  \underbrace{\sqrt{ P_{+E_+} P_{-E_+} } + \sqrt{ P_{+E_-} P_{-E_-}}}_{o_{BM}}
\eeq
one runs into contradiction with the bound derives previously (\ref{ent bound}). The separability assumption is untenable and one concludes that the mirror is entangled with the optical modes.\\ 

For the proposed device, the distance between the superposed mechanical position and  the surrounding temperature are such that $o_{BM} \sim 0.43.$ The collective coupling $g_0 \beta$ is large with respect to the mechanical frequency and the position of the mirror is accurately estimated so that $V $ is very closed to unity and $p_{11}$ is very close to zero. This should allow to detect optomechanical entanglement via the witness (\ref{witn_ent}).\\

Note that the value of $o_{BM}$ is limited by its asymptotic value of $\frac{1}{2}\sqrt{1-\left(\frac{2}{\pi}\right)^2}\sim 0.38$ reached for $g_0 \tau \beta \to \infty$. However it is possible to build up a witness less constraining by performing a more detailed experimental analysis of correlations between $B$ and $M$. In principle it is clear that finding the mirror at position $E_x =\prjct{x}_M$  prepares the state of the qubit $\rho_B^x$ in  a mixture of pure states $\ket{\psi_n}_B = \frac{1}{\sqrt{2}}\big(a_\beta^{(+)}(n) \ket{-}_B -a_\beta^{(-)}(n) \ket{+}_B \big)$ weighted by $|\bra{x}\D(  g_0 \tau n )\ket{0}|^2$
\eq
\rho_B^x =\sum_n |\bra{x}\D(  g_0 \tau \,n )\ket{0}\!|^2 \prjct{\psi_n}_B.
\eeq
 So a projective measurement on the eingenstates $\ket{\xi(x)}$ and $\ket{\xi_\perp(x)}$  of $\rho_B^x$ leads to very hight correlations, and thus a very low bound $\N_{A|BM}(\rho_{AB|M})\leq  \int dx  \sqrt{P_{\ket{\xi(x)}_B, x}P_{\ket{\xi_\perp(x)}_B, x}}=\frac{1}{\sqrt{2}} \int d  x\, p_M(x) \sqrt{1-\tr  ({\rho_B^x}^2) }\approx \frac{1}{2\sqrt{1+ g_0^2 \tau^2 \beta^2/2}}$  (valid for $\beta \gg 1$.) Here  $P_{\ket{\xi},x}= \tr\, \rho_{ABM}\, \prjct{x}_M\!\otimes\!\prjct{\xi}_B$ and $p_M(x)$ is the probability to find the mirror at position $x$. In this case, one can conclude about the presence of optomechanical entanglement as soon as
\eq
\N_{AB}> \frac{1}{2\sqrt{1+ \frac{g_0^2 \tau^2 \beta^2}{2}}}\to 0 \quad \text{for} \quad g_0 \tau \beta \to \infty
\eeq
(although accessing this bound experimentally requires an unbounded number of measurements). There is thus a tradeoff between the value of the bound and the effort invested to quantify the correlations. (The bound goes from $\N_{AB}>0.38$ for $\{E_+, E_-\}$ and $\{\prjct{+}_B, \prjct{-}_B\}$  to $\N_{AB}>0$ for continuous measurement $ \prjct{x}_M$ and $\{\prjct{\xi(x)}_B, \prjct{\xi_\perp(x)}_B\}$.)\\
\\
\\
%%%%%%%%%%%%%%%%%%
\centerline{\textit{Effect of decoherence models}}\\

The decoherence models that we study (both conventional and unconventional), operate as spatial localization, i.e. they lead to a decay of spatial coherences   
\eq
\label{coh_deco}
\dot{\coh{x}{y} }= - \underbrace{\mu \big(1-\Gamma(x-y)\big)}_{\gamma(x-y)}\coh{x}{y},
\eeq
$\mu$ describing the strength of localization (e.g. proportional to the mass) and $\Gamma$ is a function giving the characteristic length of the localization. Here we show first, that in our scenario, any localization process of this form results in a phase noise on the mode A once the position of the mirror is measured. Since any phase noise is equivalent to a weak measurement of the photon number \cite{Sekatski13}, this allows us to conclude that (i) any decoherence models (conventional or not) corresponding to a localization process (\ref{coh_deco}) can be seen as operating as a weak measurement on A. (ii) Displaced single-photon entangled state is well suited to detect their effect (even if they are weak) as it involves two components that are easily distinguishable by photon number measurements. We then quantify  the effect of such a spatial localization (\ref{coh_deco}) directly on the visibility of the interference between the two modes A and B. Specifically, we show that the sensitivity of the visibility measurement increases quadratically with the size of the optical initial state (displaced single-photon entangled state). We then apply the formula that we derive to various decoherence processes, namely conventional (environmentally induced) decoherence, and two unconventional decoherence models, quantum gravity and gravitationally induced decoherence. We conclude that from the combination of small temperature (small $T$), small mechanical dissipation (large $Q_m$) and macroscopic entanglement (large $\beta$), our proposal opens a way for testing unconventional decoherence. \\

\textit{Mirror decoherence as a weak measurement of the photon number in A.}\\

We start with a general description of the dynamics under decoherence effects. Incorporating the localization process (\ref{coh_deco}) into the dynamics of the system results in an additional term in the von Neumann equation
\eq
\dot{\rho}= - i [H, \rho] + \mu \Big( \int d\lambda (\tilde \Gamma(\lambda) -\delta(\lambda) ) e^{i \lambda \hat x} \rho\, e^{- i \lambda \hat x} \Big),
\eeq
where $\tilde \Gamma$ is the Fourrier transform of $\Gamma$ and $\delta$ is the Dirac delta-function. In the rotating frame $\bar{\rho}(t) = e^{i H t} \rho(t) e^{-i H t},$ one gets rid of the free evolution term
\eq
\dot{\bar{\rho}}(t)=\mu \Big( \int d\lambda (\tilde \Gamma(\lambda) -\delta(\lambda) ) e^{i \lambda \bar x(t)}\bar{\rho}(t)\, e^{- i \lambda \bar x(t)} \Big),
\eeq
with $\bar{x}(t)= e^{i H t}\hat  x e^{-i H t}$. This allows one to deduce the evolution of $\bar \rho$  during the infinitesimal time interval $d t$
\eq
\bar \rho (t+dt)= \L_{ \bar x(t)}^{dt}\big(\bar \rho(t) \big) = \int d \lambda f(\lambda) e^{i \lambda \bar x(t) } \bar \rho(t) e^{-i \lambda \bar x(t) },
\eeq
with $f(\lambda)=  \mu dt \tilde \Gamma (\lambda) + (1-\mu dt) \delta(\lambda).$ The final state after a time $t$ can then be found by dividing the time interval $t$ in discrete times $t_i$ separated by $dt$
\eq\label{evolution}
 e^{i H t} \rho(t) e^{-i H t}= \L_{ \bar x(t-dt)}^{d t}\circ...\circ \L_{ \bar x(0)}^{d t} \big( \rho(0) \big).
\eeq\\

For a mechanical oscillator of frequency $\w_m$, the free evolution $e^{-i \omega_m \tau \,m^\dag m }$  is a rotation in phase space leading to $\bar {x}(t)= \cos(\w_m t) \hat x + \sin(\w_m t) \hat p$, where $\hat x$ is the dimensionless operator $\hat x  =\frac{1}{\sqrt{2}}(m+m^\dag)$ and $\hat p$ its canonical conjugate. From (\ref{evolution}), we obtain
\begin{align}\label{final state}
& \bar\rho(t) = \int d\lambda_x d\lambda_p F(\lambda_x, \lambda_p) e^{i (\lambda_x \hat x + \lambda_p\hat p) }\rho(0) e^{-i( \lambda_x \hat x +\lambda_p\hat p) }\nonumber \\
&\text{with} \qquad\qquad F(\lambda_x,\lambda_p)= \\ &\nonumber 
\int\! \big(\prod_i  f(\lambda_i)d\lambda_i \big)\delta(\lambda_x\!- \!\sum_i \! C_{\w t_i}\lambda_i)\delta(\lambda_p\!-\!\sum_i\! S_{\w t_i}\lambda_i),
\end{align}
where $C_{\w t}$ and $S_{\w t}$ stand for $\cos(\w_m t)$ and $\sin(\w_m t)$.\\

We now use the previous formula to see the effect of localization processes on the optical modes A and B. In our scenario, the initial state $\rho_{ABM}(0)$ is prepared by the  interaction between the displaced mode and the mirror
\eq
\rho_{ABM}(0)= e^{i g_0 \tau \sqrt{2} \,a^\dag a \,\hat x }\rho_{AB}^i\otimes \rho_M^i e^{-i g_0 \tau \sqrt{2} \,a^\dag a \,\hat x }
\eeq
where $\rho_{AB}^i$ is the displaced single photon entangled state (eq. (1) in the main text)  and $\rho_M^i$ is the initial state of the mechanical oscillator. The system evolves freely for half a period and then at $t=\frac{\pi}{\omega_m},$ the position $\hat x$ of the mirror is measured in order to erase the information the mirror carries about the number of photons in the mode A (after half a period this information is encoded in the momentum $\hat p$ of the mirror). 
The measurement of the mirror position prepares the optical modes in the state 
\begin{align}
\rho_{AB}^f [x]= \tr_M \prjct{x}_M \rho_{ABM}(\frac{\pi}{\omega_m}) = \nonumber\\
 \tr_M \prjct{-x}_M \bar \rho_{ABM}(\frac{\pi}{\omega_m})
\end{align}
(the position of the mirror gets inverted after half a period). Plugging $\rho(0)=\rho_{ABM}(0)$  into (\ref{final state}) to obtain $\bar \rho_{ABM}(\frac{\pi}{\omega_m})$ yields an non normalized outcome state 
\begin{align}
\rho_{AB}^f[x] =\int d\lambda_p\, F(\lambda_p)\, p_M(\lambda_p-x) \times\nonumber\\e^{ig_0 \tau \sqrt{2}(\lambda_p-x)\, a^\dag a }\rho_{AB}^i e^{-i g_0 \tau \sqrt{2}(\lambda_p-x)\, a^\dag a } 
\end{align}
with $F( \lambda_p) = \int \lambda_x F(\lambda_x,\lambda_p)$ and $p_M(x) = \tr \prjct{x} \rho_M^i$. Note that the phase factor $e^{-i g_0 \tau \sqrt{2} \, a^\dag a x}$ proportional to the measurement result $x$ is removed using a feedback loop acting on the mode A (see Figure 4 in the main text). The overall final state (once we include the effect of the feedback loop and sum over all possible measurement outcomes $x$) $\rho_{AB}^f =\int dx \, e^{i g_0 \tau \sqrt{2}  x a^\dag a } \rho_{AB}^f[x] e^{-i g_0 \tau \sqrt{2} x a^\dag a }$ reads
\eq \label{result}
\rho_{AB}^f = \int d\lambda_p\, \frac{F( \frac{\lambda_p}{g_0 \tau \sqrt{2}})}{g_0 \tau \sqrt{2}}  \,e^{i  \lambda_p\, a^\dag a }\rho_{AB}^i e^{-i  \lambda_p \, a^\dag a } 
\eeq
where we rescaled the integration variable $\lambda_p \to \lambda_p/(g_0 \tau \sqrt{2})$. 
This previous formula has a direct interpretation: The entangling interaction between the mirror and the optical mode A followed by a localization process operating on the mirror and the subsequent measurement of its position after half a period has an effect on A similar to a phase noise 
\eq\label{phase noise}
\rho_{AB}^f = \int d \varphi \,\tilde \xi(\varphi)\, e^{i \varphi a^\dag a} \rho_{AB}^i  \, e^{- i \varphi a^\dag a} 
\eeq
where the phase fluctuation is governed by $\tilde\xi(\varphi) = \frac{1}{g_0 \tau \sqrt{2}} F(\frac{\varphi}{g_0 \tau \sqrt{2}})$. The exact form of $\tilde\xi$ can be determined in the following way. The equation (\ref{final state}) implies
\eq
F(\frac{\varphi}{g_0 \tau \sqrt{2}}) = \int \big(\prod_i f(\lambda_i) d\lambda_i\big) \delta(\frac{\varphi}{g_0 \tau \sqrt{2}}- \sum_i \sin(\w t_i)).
\eeq
Using the delta function representation $\delta(x) =  \frac{1}{2\pi} \int e^{i \eta x} d\eta$ gives
\eq
\tilde\xi(\varphi) = \frac{1}{2\pi} \int d\eta \,e^{i \eta \varphi}\, e^{-\mu t \big(1- \mn{ \Gamma ( \eta g_0 \tau x_0 2 \sin(\theta) ) }_\theta \big) }
\eeq
where $t = \pi/\w_m$ is half the mechanical period and 
\eq
\mn{\Gamma(\eta g_0 \tau x_0 2 \sin(\theta))}_\theta=\frac{1}{\pi}\!\int_0^{\pi}\!\!\!\!\Gamma(\eta g_0 \tau x_0 2 \sin(\theta)) d\theta
\eeq 
stands for the mean value when $\theta$ spans $[0,\pi]$. Note that the constant $\sqrt{2}\, x_0$ appearing in the previous equations make the bridge between the dimension canonical operator $\hat x$ and the position in the real space $x_m = \sqrt{2}\, x_0 \hat x$ that governs the localization scale. The correspondence between the phase noise and the weak measurement of the photon number established in \cite{Sekatski13} implies that (\ref{phase noise}) equivalently stands for a weak measurement of $a^\dag a$ by a pointer with a spatial distribution $\xi (X) = \int d\varphi \,e^{-i \varphi X} \tilde \xi (\varphi),$ i.e.
\eq
\xi(X) = e^{-\mu t \big(1- \mn{ \Gamma ( X g_0 \tau x_0 2 \sin(\theta) ) }_\theta \big) }
\eeq
i.e. 
\eq\label{xi gamma}
\xi(X) = \exp\big[\!-\! t \mn{ \gamma \Big( X g_0 \tau x_0 2 \sin(\theta)\Big) }_\theta \big]
\eeq
where $\gamma(\Delta x) = \mu(1-\Gamma(\Delta x)).$ This analogy holds for all decoherence models, conventional or not, that operate as a localization process in the form (\ref{coh_deco}). It invites us to conclude that in our scenario as in the proposals of Refs. \cite{Marshall03, Kleckner08, Pepper12prl, Pepper12}, unconventional decoherence models (such as the ones reported in Refs. \cite{Ellis84, Ellis89, Ellis92, GRW86, Pearle90, Diosi89, Penrose96}) operate in a similar way that conventional decoherence \cite{Zurek03} although their origins are deeply different. This makes them difficult to test at least at first sight. In fact, it is possible to find experimentally plausible parameters allowing one to maximize the effect of unconventional decoherence and at the same time to keep the effect of environmental standard decoherence small. This is achieved by a combination of large macroscopic distinction of the superposed mechanical center of mass states (i.e. by a large $\beta$) and small thermal decoherence (i.e. ($k_B T / \hbar Q_m$) small), as we show below. \\

\textit{Effect of mirror decoherence on the visibility of the interference between A and B.}\\

We now quantify the effect of localization processes directly on the visibility of the interference between A and B. These two modes are initally prepared in the displaced single-photon entangled state $\rho_{AB}^i$ given in Eq. (1) in the main text. Once the position of the mirror is measured, the mode A is displaced back to the single photon level through $\D_A(-\beta)$. This leads to 
\begin{align}
&\rho_{AB}^m = \D_A(-\beta) \rho_{AM}^f \D_A^\dag(-\beta) =\nonumber\\ \int d\varphi \,\tilde \xi (\varphi)\,
 &\D_A(-\beta) e^{i \varphi a^\dag a}\rho_{AB}^i e^{-i \varphi a^\dag a} \D_A(\beta).\nonumber
\end{align}
At the leading order, the measured visibility is given by $ V=2 \,|\!\bra{01}\rho_{AB}^m\ket{10}\!|.$ After straightforward algebra, we obtain
\eq
V = \Big|\int d\varphi\,\tilde \xi(\varphi) \,e^{2|\beta|^2(\cos(\varphi)-1)}(e^{i\varphi}+(1-e^{i\varphi})^2 |\beta|^2 )\Big|.
\eeq
We restrict ourself to the detection of weak localization effects. In this regime, the phase noise distribution $\tilde \xi (\varphi)$ is very narrow and the above expression can be expanded to the second order
\eq
V = 1- \left(\frac{1}{2}+ 2 \beta^2\right) \int d\varphi \, \tilde \xi(\varphi) \varphi^2
\eeq
which leads to
\eq\label{collapse on visibility}
V= 1- \left(\frac{1}{2}+2 \beta^2\right)\left( - \xi''(0)\right)
\eeq
where $\xi''(0)$ stands for the second order derivative of $\xi(X)$ estimated in $X=0.$ The expression (\ref{collapse on visibility}) clearly demonstrates that the displacement amplifies the effect of the localization by a factor proportional to $\beta^2$. Small localization effects can thus be measured on short timescales if we use displaced single-photon entangled states with macroscopically distinct components, c.f. below.\\
In the following, we use the formula (\ref{collapse on visibility}) to quantify precisely the effect of various localization processes on the observed visibility.\\

\textit{Conventional (environmentally induced) decoherence}\\

The mechanical device is coupled to a finite temperature bath which can get information about its position and ultimately, disentangle it to the optical modes. To find the timescale of this decoherence mechanism, the mechanical system can be modeled as being coupled to an infinite bath of harmonic oscillators \cite{Zurek03}. In the limit $k_b T \gg \hbar \omega_m,$ $Q_m \gg 1$, the environment can be averaged out and we end up with a master equation for the density matrix of the mechanical device which involves three terms (See for example \cite{Kleckner08} Eq. (15)). The first term represents the unitary evolution of the system, while the second term represents a damping (with the damping coefficient $\omega_m/Q_m$) and the last term a diffusion (with the diffusion coefficient $m k_b T \omega_m/Q_m$). Following Zurek \cite{Zurek03}, this master equation is dominated by the diffusion term in the macroscopic regime (highest order in $\hbar^{-1}$). By evaluating this diffusion in the position basis, we find that the localization is governed by \cite{Marshall03, Kleckner08, Pepper12}
\begin{equation}
\label{eid}
\gamma(\Delta x)= \frac{M k_b T \omega_m  \Delta x^2}{\hbar^2 Q_m}.
\end{equation}
In the main text, we estimate the timescale of environmentally induced decoherence $\gamma(\Delta x)^{-1}$ for various values of $T$ and $Q_m$ by taking the distance between the superposed mechanical positions averaged over a period $\Delta x=2g_0 \tau \beta x_0.$ The formula (\ref{collapse on visibility}) allows one to get the full evolution of the visibility as a function of the delay with which the mirror position is measured. From (\ref{xi gamma}) and (\ref{eid}), we indeed obtain
\eq
\xi_{EID}(X)= \exp\big[- t X^2 \frac{2 g_0^2 \tau^2 x_0^2\, M k_b T \omega_m}{\hbar^2 Q_m} \big]
\eeq
where $t$ is the evolution time, which is an integer multiple of half the mechanical period $t =\frac{n \pi}{\omega_M}$. Using (\ref{collapse on visibility}), we get the reduction of the visibility due to conventional (environmentally induced) decoherence
\eq
\label{V_conv}
1-V = n (1+4|\beta|^2) \frac{2 \pi  g_0^2 \tau^2 x_0^2\, M k_b T}{\hbar^2 Q_m}.
\eeq\\

\textit{Quantum gravity}\\

Quantum gravity \cite{Ellis84, Ellis89, Ellis92} is a position-localized decoherence mechanism due to coupling of the system with the spacetime. It is phenomenologically equivalent to Continuous Spontaneous Localization \cite{GRW86,Pearle90} and the corresponding master equation \cite{Pepper12} is such that
\eq
\label{gamma_quant_gravity}
\gamma(\Delta x) \approx \frac{c^4 M^2 m_0^4}{\hbar^3 m_p^3} \Delta x^2. 
\eeq
Here $m_p =\sqrt{\hbar c/G}$ is the Planck mass and $m_0$ is the nucleon mass. The timescale of quantum gravity is estimated in the main text from $\gamma(\Delta x=2g_0 \tau \beta x_0)^{-1}.$ By comparing the timescale of conventional decoherence (\ref{eid}) and quantum gravity (\ref{gamma_quant_gravity}) obtained by replacing  $\Delta x=2g_0 \tau \beta x_0$, we see that it is possible to maximize the effect of quantum gravity by choosing large macroscopic distinction of the superposed mechanical center of mass states (i.e. large $\beta$) and at the same time to keep the effect of environmental decoherence small by choosing small $T$ and large $Q_m.$ \\
Note another important feature, namely that the rate of standard decoherence scales linearly with mass, while quantum gravity scales quadratically with mass.  Therefore, in addition to absolute decoherence rates, the scaling behavior provides an independent assessment of the nature of the observed decoherence.\\

Note that in addition to the timescale over which quantum gravity operates, the formula (\ref{collapse on visibility}) allows one to get the full evolution of the visibility expected from quantum gravity. This is obtained from
\eq
\xi_{QG}(X)= \exp \big[-  X^2 \frac{n \pi}{\omega_m} \frac{2 g_0^2 \tau^2 x_0^2\,c^4 M^2 m_0^4 }{\hbar^3 m_p^3} \big]
\eeq
which leads to
\eq
\label{V_quantumgravity}
1-V = n (1+4|\beta|^2) \frac{2 \pi g_0^2 \tau^2 x_0^2\,c^4 M^2 m_0^4 }{ \w_m \hbar^3 m_p^3}
\eeq
$n$ being the number of half mechanical periods separating the optomechanical interaction and the measurement of the mirror position. Comparing the formulas (\ref{V_conv}) and (\ref{V_quantumgravity}), we conclude once more that it  is the combination of small temperature (small $T$), small mechanical dissipation (large $Q_m$) and macroscopic entanglement (large $\beta$) that allows for testing unconventional decoherence (here quantum gravity).\\
  
\textit{Gravitationally induced collapse}\\

This model which has been proposed independently by Diosi \cite{Diosi89} and Penrose \cite{Penrose96} suggests that a superposition of a massive system results in a superposition of two space-times. The failure to identify a single time structure when a local description is required may force the superposition state to collapse. To give an estimate of the corresponding decoherence time, Diosi and Penrose proposed similar formulas which use the mass distributions of the two superposed states. Although it is not clear what form of mass distributions should be taken when attempting to apply this formula, we here consider that the mass is distributed over a number of spheres corresponding to atomic nuclei (mass $m_n = Z \times m_0 \sim 10^{-25}$ kg (the atomic number Z is taken as the one of tantalum) and radius $a\sim 10^{-15}$ m). In this case, the localization is governed by \cite{Kleckner08, Pepper12}
\begin{widetext}
\begin{equation}
\gamma(\Delta x)\!=\!
\frac{8 \pi G M m_n}{\hbar}\!
\left\{\!\!\begin{array}{ll}
(\frac{6}{5a} - \frac{1}{\Delta x})& |\Delta x|>2a\\
(\frac{6}{5a}\! - \!\frac{12a^2 -5 \Delta x^2}{10 a^3}\!+\!\frac{\Delta x^5-30 a^2 \Delta x^3}{160 a^6}) &\text{else}
\end{array}\right.
\eeq
\end{widetext}
The estimation of the timescale of gravitationally induced collapse which is given in the main text is obtained by taking $\Delta x=2g_0 \tau \beta x_0$ (which is larger than $2a$.) It shows that for the proposed device, gravitationally induced collapse has a dramatic effect on the visibility as it operates in a time scale shorter than the mechanical period. The corresponding phase noise distribution is not narrow and the development that we used to get (\ref{collapse on visibility}) is not valid for the proposed parameter. More concretely, for much smaller distances (and thus narrow phase noise distribution), we would obtain
\eq
\xi''_{GIC}(0) = n\frac{\pi}{\omega_m} \frac{16\pi \,  g_0^2 \tau^2 x_0^2 \,G M m_n}{ a^3 \hbar}
\eeq
so that the effect on visibility is directly given by
\eq
\label{V_gravinduceddeco}
1-V = (\frac{1}{2}+2 |\beta|^2) \xi''_{GIC}(0)\big.
\eeq
In other words, even for $\beta$ ten times smaller than what we proposed in the main text, gravitationally induced collapse might be testable with our approach. More generally, comparing the formulas (\ref{V_conv}) and (\ref{V_gravinduceddeco}), we conclude again that it is the combination of small temperature (small $T$), small mechanical dissipation (large $Q_m$) and macroscopic entanglement (large $\beta$) that allows for testing unconventional decoherence (here gravitationally induced collapse). Note also that while the rate of standard decoherence scales linearly with mass, gravitationally induced collapse scales with (mass $\times$ the atomic number Z). Once again, in addition to absolute decoherence rates, which might be tedious to determine in an actual experiment, the scaling behavior provides an independent assessment of the nature of the observed decoherence.


\begin{thebibliography}{3}

\bibitem{GRW86} G.C. Ghirardi, A. Rimini, and T. Weber, Phys. Rev. D {\bf 34}, 470 (1986).

\bibitem{Pearle90} G.C. Ghirardi, P. Pearle, and A. Rimini, Phys. Rev. A {\bf 42}, 78 (1990).

\bibitem{Gisin89} N. Gisin, Helvetica Physica Acta {\bf 62}, 363 (1989).

\bibitem{Diosi89} L. Diosi, Phys. Rev. A {\bf 40}, 1165 (1989).

\bibitem{Penrose96} R. Penrose, Gen. Relativ. Gravit. {\bf 28}, 581 (1996).

\bibitem{Zurek03} W.H. Zurek, Rev. Mod. Phys. {\bf 75}, 715 (2003).

\bibitem{Brune96} M. Brune, E. Hagley, J. Dreyer, X. Ma\^{\i}tre, A. Maali, C. Wunderlich, J. M. Raimond, and S. Haroche, Phys. Rev. Lett. {\bf 77}, 4887 (1996).

\bibitem{Turchette00} Q. A. Turchette, C. J. Myatt, B. E. King, C. A. Sackett, D. Kielpinski, W. M. Itano, C. Monroe, and D. J. Wineland, Phys. Rev. A {\bf 62}, 053807 (2000).

\bibitem{Myatt00} C.J. Myatt, B.E. King, Q.A. Turchette, C.A. Sackett, D. Kielpinski, W. M. Itano, C. Monroe, and D. J. Wineland, Nature {\bf 403} 269 (2000).

\bibitem{Hackermueller04} L. Hackermueller, K. Hornberger, B. Brezger, A. Zeilinger, and M. Arndt, Nature {\bf 427}, 711 (2004).

\bibitem{Deleglise08} S. Deleglise, I. Dorsenko, C. Sayrin, J. Bernu, M. Brune, J.-M. Raimond, S. Haroche, Nature {\bf 455}, 510 (2008).

\bibitem{Nimmrichter11} S. Nimmrichter, P. Haslinger, K. Hornberger, and M. Arndt, New J. Phys. {\bf 13}, 075002 (2011).

\bibitem{Romero-Isart11} O. Romero-Isart, A.C. Pflanzer, F. Blaser, R. Kaltenbaek, N. Kiesel, M. Aspelmeyer, and J.I. Cirac, Phys. Rev. Lett. {\bf 107}, 020405 (2011).  

\bibitem{Oconnell10} A.D. O'Connell \textit{et al.}, Nature {\bf 464}, 697 (2010).

\bibitem{Teufel11} J.D. Teufel, T. Donner, D. Li, J.W. Harlow, M.S. Aliman, K. CIcak, A.J. Sirols, J.D. Whittakker, K.W. Lehnert, and R.W. Simmonds, Nature {\bf 475}, 359 (2011).

\bibitem{Chan11} J. Chan, T.P.M. Alegre, A.H. Safavi-Naeini, J.T. Hill, A. Krause, S. Groblacher, M. Aspelmeyer, and O. Painter, Nature {\bf 478}, 89 (2011).

\bibitem{Aspelmeyer13} See M. Aspelmeyer, T.J. Kippenberg, and F. Marquardt, arXiv:1303.0733 and references therein.

\bibitem{Bose97} S. Bose, K. Jacobs, and P.L. Knight, Phys. Rev. A {\bf 56}, 4175 (1997).

\bibitem{Bose99} S. Bose, K. Jacobs, and P.L. Knight, Phys. Rev. A {\bf 59}, 3204 (1999).

\bibitem{Marshall03} W. Marshall, C. Simon, R. Penrose, and D. Bouwmeester, Phys. Rev. Lett. {\bf 91}, 130401 (2003).

\bibitem{Kleckner08} D. Kleckner, I. Pikovski, E. Jeffrey, L. Ament, E. Eliel, J. van den Brink, and D. Bouwmeester, New J. Phys. {\bf 10}, 095020 (2008).

\bibitem{Pepper12prl} B. Pepper, R. Ghobadi, E. Jeffrey, C. Simon and D. Bouwmeester, Phys. Rev. Lett. {\bf 109}, 023601 (2012).

\bibitem{Vanner11} M.R. Vanner, I. Pikovski, G.D. Cole, M.S. Kim, C. Brukner, K. Hammerer, G.J. Milburn, and M. Aspelmeyer, Proc. Natl. Acad. Sci. U.S.A {\bf 108}, 16182 (2011).

\bibitem{Cerrillo11} S. Machnes, J. Cerrillo, M. Aspelmeyer, W. Wieczorek, M.B. Plenio, A. Retzker, Phys. Rev. Lett. {\bf 108}, 153601 (2012).

\bibitem{Wang11} X. Wang, S. Vinjanampathy, F.W. Strauch, and K. Jacobs, Phys. Rev. Lett. {\bf 107}, 177204 (2011).

\bibitem{Sekatski12} P. Sekatski, N. Sangouard, M. Stobinska, F. Bussieres, M. Afzelius, and N. Gisin, Phys. Rev. A {\bf 86}, 060301 (2012).

\bibitem{Sekatski13} P. Sekatski, N. Sangouard, and N. Gisin, arXiv:1306.0843

\bibitem{foonotesinglet} The singlet state can equivalently be written as $\frac{1}{\sqrt{2}}(|01\rangle-|10\rangle)$ and as $\frac{1}{\sqrt{2}}(|+-\rangle-|-+\rangle).$

\bibitem{Paris96} M.G.A. Paris, Phys. Lett. A {\bf 217}, 78 (1996).

\bibitem{Chou05} C.W. Chou, H. de Riedmatten, D. Felinto, S.V. Polyakov, S.J. van Enk, and H.J. Kimble, Nature {\bf 438}, 828 (2005)

\bibitem{Ellis84} J. Ellis, J.S. Hagelin, D.V. Nanopoulos, and M. Srednicki, Nucl. Phys. B {\bf 241}, 381 (1984).

\bibitem{footnote_phase_dis} Note that the same constraint also applies to the phase of the local oscillator used to displace $A$ back to the single-photon level (see \cite{Sekatski12}). 

\bibitem{footnote_phase} To obtain this result, the unitary $e^{i\frac{g_0^2 \tau^3 \omega_m}{3} \beta^4 a^{\dag} a}$ needs to be applied on $A$ between the interaction with the mechanics and the re-displacement.

\bibitem{Vanner13} M.R. Vanner, J. Hofer, G.D. Cole, and M. Aspelmeyer, Nature Communications {\bf 4}, 2295 (2013).

\bibitem{Muller10} A. Muller, E.B. Flagg, J.R. Lawall, and G.S. Solomon, Opt. Lett. {\bf 35}, 2293 (2010).

\bibitem{footnotepostselection} Considering that the detectors need to be opened for approximately $2\pi \kappa^{-1}$ and assuming detector dark count rate of about 1 Hz, delays up to 500 $\mu$s could be obtained by degrading the observed visibility by less than $1\%$ only. 

\bibitem{Pepper12} B. Pepper, E. Jeffrey, R. Ghobadi, C. Simon, and D. Bouwmeester, New J. Phys. {\bf 14}, 115025 (2012).

\bibitem{Safavi13} A. H. Safavi-Naeini, S. Groeblacher, J. T. Hill, J. Chan, M. Aspelmeyer, and O. Painter,
Nature {\bf 500}, 185 (2013).

\bibitem{Laurat07} J. Laurat \textit{et al.} Phys. Rev. Lett. {\bf 99}, 180504 (2007).

\bibitem{Choi08} K.S. Choi \textit{et al.} Nature {\bf 67} (2008).

\bibitem{Usmani12} I. Usmani \textit{et al.} Nat. Photonics {\bf 6}, 234 (2012).

\bibitem{Bruno13} N. Bruno \textit{et al.} Nat. Phys. {\bf 9}, 545 (2013).

\bibitem{Ellis89} J. Ellis, S. Mohanty, and D.V. Nanopoulos, Phys. Lett. B {\bf 221}, 113 (1989).

\bibitem{Ellis92} J. Ellis, N.E. Mavromatos, and D.V. Nanopoulos, Phys. Lett. B {\bf 293}, 37 (1992).




\end{thebibliography}
\end{document}